\documentstyle[preprint,aps,eqsecnum]{revtex}
\begin{document}
\title
{BARGMANN INVARIANTS AND GEOMETRIC PHASES -- A GENERALISED CONNECTION}
\author{Eqab M. Rabei\thanks{email: eqab@center.mutah.edu.jo}}
\address{ Department of Physics, Mutah University, PostBox 7, Karak, Jordan} 
\author{Arvind\thanks{email: arvind@physics.iisc.ernet.in}}
\address{ Department of Physics, Guru Nanak Dev University, Amritsar 143005, 
India}
\author{N. Mukunda\thanks{email: nmukunda@cts.iisc.ernet.in} \thanks{Honorary 
Professor, Jawaharlal Nehru Centre for
Advanced Scientific Research, Jakkur, Bangalore 560064}}
\address{Centre for Theoretical Studies and Department  of Physics, 
Indian Institute of Science,
Bangalore 560012, India}
\author{R. Simon\thanks{email: simon@imsc.ernet.in}}
\address{ The Institute of Mathematical 
Sciences, C.I.T. Campus, Chennai 600113, 
India}
\date{\today}
\maketitle
\begin{abstract}
We develop the widest possible generalisation of the well-known connection between 
quantum mechanical Bargmann invariants and geometric phases.  The key notion is that 
of null phase curves in quantum mechanical ray and Hilbert spaces.  Examples of such 
curves are developed.  Our generalisation is shown to be essential to properly 
understand geometric phase results in the cases of coherent states and of Gaussian 
states. Differential geometric aspects of null phase curves are also briefly explored.
\end{abstract} 
\section{Introduction}
The geometric phase was originally discovered in the context of cyclic adiabatic 
quantum mechanical evolution, governed by the time-dependent Schr\"{o}dinger equation 
with a hermitian Hamiltonian operator\cite{berry,reprint}.  Subsequent work has shown 
that many of 
these restrictions can be removed.  Thus the geometric phase can be defined in 
nonadiabatic\cite{anandan}, noncyclic and even nonunitary evolution\cite{samuel}.  
Generalization to the nonabelian case has also been achieved\cite{nonabelian}. 
Finally the kinematic 
approach\cite{kinematic} demonstrated that even the Schr\"{o}dinger equation and a 
Hamiltonian operator are not needed for defining the geometric phase. The intimate 
relationship between geometric phase and Hamilton's theory of turns\cite{hamilton} has
 also been brought out\cite{jphysa}.

An important consequence of the kinematic approach has been to bring out clearly the 
close connection between geometric phases, and a family of quantum mechanical 
invariants introduced by Bargmann\cite{bargmann} 
 while giving a new proof of the Wigner\cite{wigner} 
unitary-antiunitary
 theorem.  This connection depends in an essential  way upon the concept of free 
geodesics in quantum mechanical ray and Hilbert spaces, and the vanishing of geometric 
phases for these geodesics.

The purpose of this paper is to generalise this important link between  Bargmann  
invariants and geometric phases to the widest possible extent, by going beyond the 
use of free geodesics.  The key is to characterize in a complete way those ray space 
curves with the property that the geometric phase vanishes for any connected stretch 
of any one of them.  We show that this property can be translated into an elementary 
and elegant statement concerning the inner product of any two Hilbert space vectors 
along any lift of such a ray space curve.  We refer to these as ``null phase curves''; 
and the generalisation of the familiar statement linking Bargmann invariants and 
geometric phases is achieved by replacing free geodesics by such curves.  A free 
geodesic is always a null phase curve; however the latter is a much 
more general object.

The material of this paper is arranged as follows.  Section 2 recalls the basic 
features of the kinematic approach to the geometric phase; sets up free geodesics in 
ray and Hilbert spaces; shows that the geometric phase for any free geodesic vanishes; 
introduces the Bargmann invariants; and describes their connection to geometric phases 
for ray space polygons bounded by free geodesics.  In Section 3 it is argued that it 
should be possible to generalise this connection.  This motivates the definition 
and complete characterization of null phase curves at the Hilbert space level, the 
previous free geodesics being a very particular case.  
It is then shown that such curves allow us to generalise the previously stated 
connection to the widest possible extent.  Section 4 defines the concept of 
constrained geodesics in ray and Hilbert spaces, the motivation being that in some 
situations such curves may in fact be null phase curves.  The idea is extremely 
simple, namely we limit ourselves to some chosen submanifolds in ray (and Hilbert) 
space, and determine curves of minimum length lying within these submanifolds.  
Section 5 examines several interesting examples to illustrate these ideas: a 
submanifold arising out of a linear subspace of Hilbert space; coherent states 
for one degree of freedom; centered Gaussian pure states 
for one degree of freedom; and an 
interesting submanifold in the space of two-mode coherent states.   
It turns out that in the first case constrained geodesics are just free geodesics, 
while in the remaining cases they are very different.  This shows that the 
generalised connection between Bargmann invariants and geometric phases enunciated 
in this paper is just what is needed to be physically interesting and appropriate.  
In Section 6 we present a brief discussion of these ideas in the differential 
geometric framework natural to geometric phases, and also develop a direct ray 
space description of null phase curves; while Section 7  contains concluding remarks.
\section{The connection based on free geodesics}
Let $ {\cal H}$ be the Hilbert space of states of some quantum 
system, ${\cal R}$ the associated ray space, 
and $\pi : {\cal H}\rightarrow {\cal R}$ the corresponding projection.  
We shall be dealing with (sufficiently) smooth parametrised curves ${\cal C}$ of 
unit vectors in ${\cal H}$, and their images $C$ in ${\cal R}$.  
 A curve ${\cal C}$ is described as follows:
     \begin{eqnarray}
     {\cal C}=\left\{\psi(s)\in{\cal H}\;\;\big |\;\;\parallel \!\psi(s) 
\!\parallel\; =\, 1\,,\;\;\; s_1\leq
     s\leq s_2\right\} \subset {\cal H}\,.
     \end{eqnarray}

\noindent
Its image $C$ is a curve of pure state density matrices:
     \begin{eqnarray}
     \pi[{\cal C}] &=& C \subset {\cal R}\;,\nonumber\\
     C &=& \left\{\rho(s) = \psi(s) \psi(s)^{\dagger} \;\;\big |\;\; s_1 \leq s 
\leq            s_2\right\}\,.
     \end{eqnarray}

\noindent
Any ${\cal C}$ in ${\cal H}$ projecting on to a given $C$ in ${\cal R}$ is a lift 
of the latter.  In particular we have a horizontal lift ${\cal C}^{(h)}$ if the 
  vectors $\psi^{(h)}(s)$ along it are such that
     \begin{eqnarray}
     \left(\psi^{(h)}(s), \frac{d}{ds} \psi^{(h)}(s)\right) = 0\,.
     \end{eqnarray}

For any curve $C\subset {\cal R}$ a geometric phase $\varphi_g[C]$ is defined.  
Its calculation is facilitated by going to any lift ${\cal C}$, calculating the 
total and dynamical phases for ${\cal C}$, and taking the difference: 
     \begin{eqnarray}
     \pi[{\cal C}] = C:\;\;\;\;&&\nonumber\\
     \varphi_g[C] &=& \varphi_{{\rm tot}} [{\cal C}] - \varphi_{\rm dyn}
     [{\cal C}]\,,\nonumber\\
     \varphi_{\rm tot}[{\cal C}] &=& \arg (\psi(s_1), \psi(s_2))\,,\nonumber\\
      \varphi_{\rm dyn}[{\cal C}] &=&
      \mbox{Im} \int\limits^{s_{2}}_{s_{1}} ds \left(\psi(s),
      \frac{d}{ds} \psi(s)\right)\,.
      \end{eqnarray}

\noindent
In particular, if ${\cal C}$ is horizontal 
$\varphi_{\rm dyn}\protect[{\cal C}\protect]$ vanishes, and $\varphi_g[C]$ 
is just $\varphi_{\rm tot}\protect[{\cal C}\protect]$.

Now we define free geodesics in ${\cal R}$ and ${\cal H}$.  Given $C$ in ${\cal R}$ 
and any lift ${\cal C}$ in ${\cal H}$, the length of the former can be defined as the 
following nondegenerate functional:
     \begin{eqnarray}
     L[C] = \int\limits^{s_{2}}_{s_{1}} ds     
      \left\{\big | \big |\frac{d\psi(s)}{ds}\big |\big |^2 - \big| \left(\psi(s),   
   \frac{d\psi(s)}{ds}\right)\big| ^2\right\}^{1/2}\,.
     \end{eqnarray}

\noindent
It is easy to check that the integrand here is independent of the choice of 
lift ${\cal C}$; it leads to the well known Fubini-Study metric on ${\cal 
R}$\cite{page,kobayashi}.   Free geodesics in ${\cal R}$ are those $C$'s for which 
$L[C]$ 
is a minimum for given end points.  And by definition a free geodesic in ${\cal H}$ 
is any lift of a 
 free geodesic in ${\cal R}$.  It can be shown\cite{kinematic} that any free 
geodesic in ${\cal 
R}$ can be lifted to ${\cal H}$, and the parametrisation chosen, so that it can 
be described as follows:
     \begin{eqnarray}
     \psi(s) &=& \phi _1 \cos s + \phi _2 \sin  s\,,\nonumber\\
     (\phi_1, \phi_1) &=& (\phi_2, \phi_2) = 1,\;\;\;\; (\phi_1, \phi_2) = 0\,.
      \end{eqnarray}

\noindent
Thus we have here a plane two-dimensional curve determined by a pair of orthonormal 
vectors in ${\cal H}$, an arc of a circle.  
It may be helpful to make the following comment 
concerning free geodesics.  Given any two `non-orthogonal' points 
$\rho_1,\rho_2\in {\cal R}$, that is such that $\mbox{Tr}(\rho_1 \rho_2)\neq 0$, 
we can always choose unit vectors $\psi_1, \psi_2\in {\cal H}$ projecting 
onto $\rho_1,\rho_2$ respectively, such that the inner product $(\psi_1,\psi_2)$ 
is real positive.  Then the free geodesic (2.6) will connect $\psi_1$ and $\psi_2$ 
if we take $\phi_1=\psi_1$ and $\phi_2=(\psi_2
-\psi_1(\psi_1,\psi_2))/\left\{1 - (\psi_1,\psi_2)^2\right\}^{1/2}$.  
It is now clear that $\psi(0)=\psi_1$, and $\psi(s)=\psi_2$ 
for $s=\cos^{-1}(\psi_1,\psi_2) \in (0,\pi/2)$.
It is clear that the curve in ${\cal H}$ 
given by eqn. (2.6) is horizontal; and for any two  points on 
it with $\vert s_1-s_2\vert < \pi/2\/$,
the inner product $(\psi(s_1), \psi(s_2))$ is real positive, so $\psi(s_1)$ 
and $\psi(s_2)$ are in phase in the Pancharatnam sense\cite{pancharatnam}. From these 
properties of free geodesics the result\cite{kinematic}
     \begin{eqnarray}
     \varphi_g[\mbox{free geodesic in}\;{\cal R}] = 0
      \end{eqnarray}

\noindent
follows.  This can be exploited to connect geometric phases to Bargmann invariants.  

Let $\psi_1, \psi_2,\ldots, \psi_n$ be any $n$ unit vectors in ${\cal H}$, no two 
consecutive ones being orthogonal; and let $\rho_1,\rho_2,\ldots,\rho_n$ be their 
images in ${\cal R}$.  Then the corresponding $n$-vertex Bargmann invariant is 
defined as
     \begin{eqnarray}
     \Delta_n(\psi_1,\psi_2,\ldots,\psi_n)
      &=& (\psi_1,\psi_2)(\psi_2,\psi_3)\ldots (\psi_n,\psi_1)\nonumber\\
      &=&\mbox{Tr} (\rho_1 \rho_2\ldots \rho_n)\,.
      \end{eqnarray}

\noindent
Now we draw $n$ free geodesics in ${\cal R}$ connecting $\rho_1$ 
to $\rho_2$, $\rho_2$ to $\rho_3, \ldots, \rho_n$ to $\rho_1$.  
Thus we obtain an $n$-sided polygon in ${\cal R}$ bounded by free 
geodesics, and we can compute the corresponding geometric phase.  
Repeatedly exploiting eqn.(2.7) we obtain the basic result\cite{kinematic}:
     \begin{eqnarray}
     \varphi_g\left[\begin{array}{c}n\mbox{-vertex polygon in}\;{\cal 
R}\;\mbox{connecting}\;
     \rho_1\;\mbox{to}\;\rho_2, \\
      \rho_2\;\mbox{to}\;\rho_3,\ldots,
     \rho_n\;\mbox{to}\;\rho_1\;\mbox{by free geodesics}\end{array}\right]
       &=& - \arg \Delta_n (\psi_1,\psi_2,\ldots,\psi_n)\,,\nonumber\\
       \rho_j&=&\psi_j\psi_j^{\dag}\,,\;\;\;j=1,2,\ldots, n\,. 
       \end{eqnarray}

\noindent
We mention in passing that this result is of considerable conceptual as well as 
practical value\cite{khanna}.

In connection with the above result, the following remarks may be made. As is clear
from equation~(2.8), the phases of the individual vectors $\psi_1,\psi_2 \cdots \psi_n\/$
can be freely altered. We need only to assume that successive pairs of unit vectors
are not mutually orthogonal; then the Bargmann invariant is non zero and has a well
defined phase.
\section{A generalised connection}
The definition (2.8) of the Bargmann invariant requires only the choice of the $n$ 
vertices $\rho_1,\rho_2,\ldots,\rho_n\in{\cal R}$; consecutive ones need not be 
connected in any way to form a closed figure.  This suggests that the connection 
(2.9) between these invariants and geometric phases may hold more generally, not 
only in the case where we connect $\rho_1$ to $\rho_2,\rho_2$ to $\rho_3,\ldots,\rho_n$
 to $\rho_1$ by free geodesics.  We now show that this is indeed so.

We need to characterize the most general (smooth) curves $C\subset {\cal R}$ having 
the property
     \begin{eqnarray}
     \varphi_g[\mbox{any connected portion of}\; C]=0\,.
      \end{eqnarray}

\noindent
We know that if $C$ is a free geodesic, this property does follow; but there may 
be (indeed there are) many other possibilities.  We can develop a simple necessary 
and sufficient condition on $C$ such that (3.1) holds.

Given the curve $C\subset {\cal R}$, let ${\cal C}^{(h)}$ be a horizontal 
lift and ${\cal C}$ a general lift of $C$ in ${\cal H}$.  We have
     \begin{eqnarray}
     C&=& \left\{\rho(s)\;\;\big|\;\;\rho(s)^{\dag}=\rho(s)\geq 0\,,\;\;\; \rho(s)^2
      = \rho(s)\,,\;\;\; \mbox{Tr}\;\rho(s)=1\,,\;\;\; s_1\leq s\leq 
s_2\right\};\nonumber\\
      {\cal C}^{(h)}&=& \left\{\psi^{(h)}(s)\in {\cal H}\;\;\big|\;\;\pi
       \left(\psi^{(h)}(s)\right)=\rho(s)\,,\;\;\; \left(\psi^{(h)}(s), \frac{d}{ds}
       \psi^{(h)}(s)\right)=0\right\};\nonumber\\
       {\cal C}&=& \left\{\psi(s)\in{\cal H}\;\;\big|\;\;\psi(s) = e^{i\alpha(s)}
       \psi^{(h)}(s)\right\}\,.
       \end{eqnarray}

\noindent
Here $\alpha(s)$ is some (smoothly varying) phase angle.  For any two points on $C$ 
with parameter values $s$ and $s^{\prime}> s$ we have:
     \begin{eqnarray}
     \varphi_g[\rho(s)\;\mbox{to}\;\rho(s^{\prime})\;
     \mbox{along}\;C]
     &=& \varphi_{\rm tot}\left\protect[\psi^{(h)}(s)\;\mbox{to}\;
     \psi^{(h)}(s^{\prime})\;\mbox{along}\; {\cal      C}^{(h)}\right\protect]\nonumber\\
     &=&\arg \left(\psi^{(h)}(s), \psi^{(h)}(s^{\prime})\right)\nonumber\\
     &=&\arg \left(e^{-i\alpha(s)}\psi(s), 
            e^{- i\alpha(s^{\prime})}\psi(s^{\prime})\right)\nonumber\\
     &=&\arg(\psi(s), \psi(s^{\prime})) + \alpha(s)-\alpha(s^{\prime})\,.
     \end{eqnarray}

\noindent
From this result we see that the necessary and sufficient condition on $C$ to secure 
the property (3.1) can be expressed in several equivalent ways, using either an 
arbitrary lift ${\cal C}$ of $C$ or a horizontal lift ${\cal C}^{(h)}$:
     \begin{eqnarray}
     \varphi_g[\mbox{any connected portion of}\;C] &=& 0       
      \nonumber\\
  \Longleftrightarrow&& \arg(\psi(s),\psi(s^{\prime})) = \alpha(s^{\prime}) - \alpha(s),
   \;\; \mbox{any}\; s^{\prime}\;\mbox{and}\;s \nonumber\\
 \Longleftrightarrow&&  \frac{\partial^2}{\partial s^{\prime}\partial s}
     \arg(\psi(s), \psi(s^{\prime})) = 0 \nonumber\\
 \Longleftrightarrow&&  \arg (\psi(s),\psi(s^{\prime})) = \mbox{separable in}\;s^{\prime}\;
     \mbox{and}\; s  \nonumber\\
     \Longleftrightarrow&& \left(\psi^{(h)}(s), \psi^{(h)}(s^{\prime})\right) =  
            \mbox{real  positive},\;\;\mbox{ any}\;
      s^{\prime}\;\mbox{and}\; s \nonumber\\
\Longleftrightarrow&& \mbox{any two points of }\; {\cal C}^{(h)} \;\mbox{ are in 
phase}\,.
     \end{eqnarray}

Here separability is to be understood in the additive, and not in the multiplicative, 
sense.  It is important to recognise that these characterizations are reparametrization
  invariant.  Any curve $C\subset {\cal R}$ 
obeying (3.4) will be called a ``null phase curve in ${\cal R}$''; and any lift 
${\cal C}$ of such a $C$ will be called a ``null phase curve in ${\cal H}$''.  
Free geodesics
 are null phase curves, but not necessarily conversely.

It may be helpful to make some additional remarks at this point to clarify the ideas 
involved.  If a curve ${\cal C}\subset {\cal H}$ is such that any two points on 
it (not too far apart) are in phase, then it is definitely horizontal:
     \begin{eqnarray}
     {\cal C} = \{\psi(s)\}:\;\; (\psi(s), \psi(s^{\prime})) &=&
      \mbox{real positive}\;\nonumber\\
  &\Longrightarrow& \left(\psi(s), \frac{d\psi(s^{\prime})}{ds^{\prime}}\right)
      = \mbox{real}\; \nonumber\\
   &\Longrightarrow& \left(\psi(s), \frac{d\psi(s)}{ds}\right) = 0 \nonumber\\
   &\Longrightarrow&       {\cal C}\;\mbox{horizontal}\,.
      \end{eqnarray}

\noindent
The image $C=\pi[{\cal C}]$ is obviously a null phase curve in ${\cal R}$ as eqn.(3.4) 
is obeyed with $\alpha(s)=0$; therefore ${\cal C}$ being a lift of $C$ is also a 
null phase curve in ${\cal H}$.  On the other hand, for a horizontal curve 
${\cal C}^{(h)} \subset {\cal H}$ only ``nearby points'' are in phase:
     \begin{eqnarray}
     {\cal C}^{(h)} =  \left\{\psi^{(h)}(s)\right\} =
      \mbox{horizontal}&& \nonumber\\
     \Longrightarrow&& \left(\psi^{(h)}(s), \frac{d}{ds} \psi^{(h)}(s)\right) = 0 
     \nonumber\\
     \Longrightarrow&& \left(\psi^{(h)}(s), \psi^{(h)}(s+\delta s)\right) \simeq 
       1+0(\delta      s)^2 \nonumber\\
     \Longrightarrow&& \arg\left(\psi^{(h)}(s), \psi^{(h)}(s+\delta s)\right) = 
      0(\delta s)^2\,.
     \end{eqnarray}

\noindent
However two general points on ${\cal C}^{(h)}$ may well be not in phase, as 
$\arg(\psi(s),\psi(s'))$ could be nonzero.  Hence ${\cal C}^{(h)}$ and its image 
$\pi[{\cal C}^{(h)}]$ may not be null phase curves.  For $\pi[{\cal C}^{(h)}]$ to 
be a null phase curve, in addition to being horizontal (a local property) 
${\cal C}^{(h)}$ must possess the global property that for general $s$ and 
$s^{\prime}$ the inner product $\left(\psi^{(h)}(s),\psi^{(h)}(s^{\prime})\right)$ 
is real positive.  This is what is captured in the conditions (3.4).

We can now generalise the result (2.9) and strengthen it as follows.  Given $n$ 
unit vectors $\psi_1, \psi_2,\ldots, \psi_n\in {\cal H}$ with images 
$\rho_1, \rho_2,\ldots, \rho_n \in {\cal R}$ : draw any null phase curves 
joining consecutive pairs of points 
$\rho_1$ to $\rho_2,\rho_2$ to $\rho_3, \ldots, \rho_n$ to $\rho_1$.  
(This can certainly be done since in any event free geodesics are available).  
Then by exactly the same arguments that lead to the connection (2.9) we obtain:
     \begin{eqnarray}
     \varphi_g\left[\begin{array}{c}n \mbox{-sided figure in}\;{\cal R}\; \mbox{with 
vertices}\;
      \rho_1,\rho_2,\ldots,\rho_n \\
      \mbox{  and bounded by null phase curves}\end{array}\right] = -\arg \Delta_n 
      (\psi_1, \psi_2, \ldots, \psi_n)\,.
       \end{eqnarray}

\noindent
It must be clear that this is the widest generalisation of the connection (2.9) 
that one can obtain.  We see that we can replace each free geodesic belonging to 
a polygon in ${\cal R}$ by any null phase curve, and the geometric phase remains  
the same, since the right hand side of eqn.(3.7) depends on the vertices alone.
\section{Constrained geodesics as null phase curves}
We have seen that every free geodesic is a null phase curve, but the converse is 
generally not true.  Nevertheless the former fact motivates the following: can we 
alter the definition of a free geodesic, based on minimising the length functional 
$L[C]$ of eqn.(2.5), in a natural way to obtain other kinds of geodesics, and will 
they turn out to be null phase curves as well?

The generalisation we explore is the following: instead of dealing with curves 
(of unit vectors) in the complete Hilbert and ray spaces ${\cal H}$ and ${\cal R}$, 
we restrict ourselves to some (smooth) submanifold $M\subset {\cal R}$ and consider 
only curves $C$ lying in $M$ and connecting pairs of points in $M$.  For such curves 
we minimise $L[C]$ with respect to variations of $C$ which stay within $M$.  The 
resulting curves will naturally be called ``constrained geodesics'', and the question 
is: do constrained geodesics in some cases turn out to  be null phase curves?

We emphasize that our question is not whether every null phase curve is a constrained 
geodesic lying in a suitably chosen submanifold $M\subset {\cal R}$, but rather whether
 the latter curves sometimes have the former property.  The physically important 
examples presented in the next Section show that our question is indeed interesting.  
In this Section we set up the general framework to handle constrained geodesics in ray 
space.  

Given ${\cal H}$ and ${\cal R}$ with $\dim {\cal H} = \dim {\cal R}+1$ in the real 
sense, we consider a submanifold $M\subset {\cal R}$ of $n$ (real) dimensions 
consisting of a (sufficiently smooth) family of unit rays, with (local) real 
independent and 
 essential coordinates $\xi=(\xi^{\mu}), \mu=1,2,\ldots,n$:
     \begin{eqnarray}
     M = \left\{\rho(\xi)\;\in\; {\cal R} \;\;\big|\;\; \xi\;\in\; 
       \Re^n\right\}\subset 
     {\cal R} \,.
     \end{eqnarray}

\noindent
(We do not indicate explicitly the domain in $\Re^n$ over which $\xi$ may vary).  
The inverse image of $M$ in ${\cal H}$ will bring in an extra phase angle $\alpha$, 
and is denoted by ${\cal M}$:
     \begin{eqnarray}
     {\cal M} &=& \pi^{-1}[M]\nonumber\\
     &=&\left \{\psi(\xi;\alpha)\;\in\; {\cal H} \;\;\big|\;\; \pi(\psi(\xi;\alpha)) =
     \rho(\xi)\,,\;\;\; \psi(\xi;\alpha) = e^{i\alpha} \psi(\xi;0)\right\}\,.
     \end{eqnarray}

\noindent
(Of course each $\psi(\xi;\alpha)$ is a unit vector, and $\alpha$ and $\xi^\mu$ 
taken together are local coordinates for ${\cal M}$ ).  So in the real sense 
$\dim {\cal M}=n+1$, and to avoid trivialities we must have $1+n/2 <$ complex 
dimension of ${\cal H}$.

Now we consider a parametrised curve $C\subset M\subset {\cal R}$, obtained by making 
the $n$ real variables $\xi^{\mu}$ into functions of a real parameter $s$:
     \begin{eqnarray}
      C = \{\rho (\xi(s))\,,\;\;\; s_1\leq s \leq s_2\}\subset M\,.
      \end{eqnarray}

\noindent
To lift $C$ to some ${\cal C}\subset{\cal M}\subset {\cal H}$, some (smooth) choice 
of phase angle $\alpha(s)$ as a function of $s$ must be made, and then we have:
     \begin{eqnarray}
     {\cal C}&=&\{\Psi(s) = \psi(\xi(s); \alpha(s))\}\subset {\cal M}\;,
     \nonumber\\
      \pi [{\cal C}]&=& C\,.
      \end{eqnarray}

\noindent
Using the definition (2.5) the length $L[C]$ can be seen to involve only the partial 
derivatives of $\psi(\xi;\alpha)$ with respect to the $\xi^{\mu}$, the dependence 
on $\alpha$ being trivial and not contributing at all.  Therefore we define:
     \begin{eqnarray}
      u_{\mu}(\xi;\alpha) &=& \frac{\partial}{\partial
                           \xi^{\mu}} \psi(\xi;\alpha)\,,\;\;\; \mu=1,2,\ldots, 
n\,;\nonumber\\
      u_{\mu}^{\perp}(\xi;\alpha) &=& u_{\mu}(\xi;\alpha) -
       \psi(\xi;\alpha) (\psi(\xi;\alpha),u_{\mu}(\xi;\alpha))\,.
      \end{eqnarray}

\noindent
Normalisation of $\psi(\xi;\alpha)$ to unity for all $\xi$ and $\alpha$ implies
     \begin{eqnarray}
      \mbox{Re}\left(\psi(\xi;\alpha), u_{\mu}(\xi;\alpha)\right)=0\,.
      \end{eqnarray}

\noindent
Now $L[C]$ can be expressed as follows:
     \begin{eqnarray}
     L[C]&=& \int\limits^{s_{2}}_{s_{1}} \!ds\sqrt{\parallel\!                 
\dot{\Psi}(s)\!\parallel^2 - |(\Psi(s), \dot{\Psi}(s))|^2\;}
     \nonumber\\
     &=&\int\limits^{s_{2}}_{s_{1}}\! ds \sqrt{g_{\mu\nu}(\xi) \dot{\xi}
      ^{\mu}\dot{\xi}^{\nu}\;}\,,\nonumber\\
      g_{\mu\nu}(\xi)&=&\mbox{Re}\left(u^{\perp}_{\mu} (\xi;\alpha),
      u^{\perp}_{\nu}(\xi;\alpha)\right)\;,\nonumber\\
      \xi^{\mu}&=& \xi^{\mu}(s)\,.
      \end{eqnarray}

\noindent
The parameter dependences of $\xi$ and $\alpha$ are as in eqn.(4.4).  
From the essentiality of $\xi^{\mu}$ as coordinates for $M$, and the positivity 
of the metric on ${\cal H}$, one easily obtains the following results: the $n\times n$ 
matrix 
$\left(\left(u_{\mu}^{\perp}(\xi;\alpha), u^{\perp}_{\nu}(\xi;\alpha)\right)\right)$ 
is hermitian positive definite and independent of $\alpha$; and only its real part 
$(g_{\mu\nu}(\xi))$, which is symmetric positive definite, enters $L[C]$.

To obtain the differential equations for constrained geodesics we minimise 
$L[C]$ with respect to variations in $C$ that stay within $M$.  This amounts 
to minimising $L[C]$ in the final form given in eqn.(4.7), by making independent 
variations in the $n$ 
real functions $\xi^{\mu}(s)$; and the result is well-known from Riemannian 
geometry.  After making a suitable choice of the parameter $s$ 
(affine parametrisation), the differential equations for constrained geodesics become:
     \begin{eqnarray}
     \ddot{\xi}^{\mu}(s) &+& \Gamma^{\mu}\,_{\nu\lambda}(\xi(s))    
     \dot{\xi}^{\nu}(s) \dot{\xi}^{\lambda}(s) = 0\;,\nonumber\\
     \Gamma^{\mu}\,_{\nu\lambda}(\xi)&=&\frac{1}{2} g^{\mu\rho}(\xi) 
     (g_{\rho\nu,\lambda}(\xi) + g_{\rho\lambda,\nu}(\xi)-
     g_{\nu\lambda,\rho}(\xi))\;,\nonumber\\
     (g^{\mu\nu}(\xi))&=& (g_{\mu\nu}(\xi))^{-1}\;,\nonumber\\
     g_{\rho\nu,\lambda}(\xi) &=& \frac{\partial g_{\rho\nu}(\xi)}
     {\partial \xi^{\lambda}}\,.
     \end{eqnarray}

\noindent
Here the $\Gamma$'s are the familiar symmetric Christoffel symbols determined by 
the `metric' tensor $g_{\mu\nu}(\xi)$.  The remaining freedom in the choice of 
parameter $s$ is only change in scale and shift of origin.  It is a consequence 
of the differential equations above that
     \begin{eqnarray}
     g_{\mu\nu}(\xi(s)) \dot{\xi}^{\mu}(s) \dot{\xi}^{\nu}(s) =
     \mbox{constant}\,.
     \end{eqnarray}

A general solution to eqn.(4.8) is uniquely determined by choices of initial 
values $\xi^{\mu}(0), \dot{\xi}^{\mu}(0)$.  The resulting $\xi^{\mu}(s)$ determine 
some constrained geodesic $C\subset M\subset {\cal R}$; and for any (smooth) 
choice of $\alpha(s)$ we get a lift ${\cal C}\subset {\cal M}\subset{\cal H}$ 
which by definition is a constrained geodesic in ${\cal H}$.  The meaning of 
the `conservation law' (4.9) in terms of Hilbert space vectors is interesting.  
In terms of the derivative of $\Psi(s)$ with respect to $s$, and its component 
orthogonal to $\Psi(s)$,
     \begin{eqnarray}
      \dot{\Psi}(s)&=& \frac{d}{ds} \psi(\xi(s); \alpha(s))\nonumber\\
      &=& \dot{\xi}^{\mu}(s) u_{\mu} (\xi(s); \alpha(s)) +i\,\dot{\alpha}(s)
       \Psi(s)\,,\nonumber\\
      \dot{\Psi}^{\perp}(s)&=& \dot{\Psi}(s) -\Psi(s) (\Psi(s),
       \dot{\Psi}(s))\nonumber\\
       &=& \dot{\xi}^{\mu}(s) u^{\perp}_{\mu} (\xi(s); \alpha(s))\,,
       \end{eqnarray}

\noindent
we have,
     \begin{eqnarray}
     g_{\mu\nu}(\xi(s)) \dot{\xi}^{\mu}(s) \dot{\xi}^{\nu}(s) &=&
     \mbox{constant}\nonumber\\
     &\Longrightarrow&\parallel \!\dot{\Psi}^{\perp}(s)\!\parallel = 
\mbox{constant}\,.
     \end{eqnarray}

\noindent
We can then, if we wish, adjust the scale of $s$ so that $\dot{\Psi}^{\perp}$ becomes 
a unit vector for all $s$.

Having set up the basic formalism to determine constrained geodesics, in the next 
Section we look at some physically motivated examples to see whether they are 
sometimes  null phase curves as well.
\section{Applications}
We look at four examples to illustrate the use of constrained geodesics  in the 
geometric phase context, and to show the distinction in general between them and 
null phase curves.

\noindent
(a) {\bf  Subspaces of } ${\cal H}$\\
Let ${\cal H}_0$ be a linear subspace of ${\cal H}$ (as a complex vector space), 
and denote by ${\cal M}\subset {\cal H}_0$ the subset of unit vectors in ${\cal H}_0$. 
 By projection we obtain the submanifold  $M=\pi[{\cal M}]\subset {\cal R}$, with 
the real dimension of $M$ equal to 
$2\times\{(\mbox{complex dimension of}\;{\cal H}_0)-1\}$.  In this case 
constrained geodesics in $M$ happen to be free geodesics.  For, given any 
two (non orthogonal) in phase unit vectors in ${\cal M}$, say $\psi_1$ and 
$\psi_2$, the free geodesic connecting them, namely from eqn.(2.6) the curve 
${\cal C}$ consisting of the vectors
     \begin{eqnarray}
     \psi(s) = \psi_1\cos s + \frac{(\psi_{2}- 
                (\psi_{1},\psi_{2})\psi_1)}{\sqrt{1- (\psi_1,\psi_2)^2}}\,
      \sin s\,,
      \end{eqnarray}

\noindent
passes entirely through points of ${\cal M}$.  Hence its image 
$\pi[{\cal C}]=C$ lies entirely within $M$; and being the free 
geodesic connecting $\pi(\psi_1)$ to $\pi(\psi_2)$ it must be the 
constrained geodesic as well.  In this case therefore we do not
 get anything new.

Conversely we see that to have a situation where constrained geodesics are different 
from free ones, the submanifold $M\subset {\cal R}$ must {\em not} arise from 
a subspace of ${\cal H}$ in the above manner.  We now look at two such cases, of 
obvious physical importance, in which true generalisations of the original Bargmann 
invariant-geometric phase connection appear.

\noindent
(b) {\bf  Single mode coherent states}

We consider the family of coherent states for a single degree of freedom, described 
by hermitian operators $\hat{q},\hat{p}$ or the nonhermitian combinations 
$\hat{a},\hat{a}^{\dag}$ :
     \begin{eqnarray}
     \hat{a}=\frac{1}{\sqrt{2}}(\hat{q} + i\,\hat{p})\,,\;\;&&\;\;                 
\hat{a}^{\dag}=\frac{1}{\sqrt{2}} (\hat{q}-i\,\hat{p})\;,\nonumber\\
      \protect[\hat{q},\hat{p}\protect] = i\,,\;\;&&\;\; \protect[\hat{a}, 
\hat{a}^{\dag}\protect] = 1\,.
      \end{eqnarray}

\noindent
A general normalised coherent state is labelled by a complex number $z$ and is 
generated by applying a unitary phase space displacement operator to the (Fock) 
vacuum state $|0\rangle$:
     \begin{eqnarray}
     |z \rangle  &=& \exp (z\hat{a}^{\dag}-z^{*}\hat{a})\, |0\rangle \nonumber\\
     &=& \exp \left(-\frac{1}{2} z^{*} z +  z\hat{a}^{\dag}\right) |0 \rangle \;,
      \nonumber\\
      \hat{a} | z\rangle  &=& z|z \rangle \,.
      \end{eqnarray}

\noindent
To conform to the notations of the previous Section, we introduce real parameters 
$\xi_1,\xi_2$, include a phase angle $\alpha$, and express the above states in terms 
of $\hat{q}$ and $\hat{p}$ as follows (for ease in writing we use $\xi_{1,2}$ rather 
than $\xi^{1,2}$):
     \begin{eqnarray}
     z = \frac{1}{\sqrt{2}} (\xi_1 +i\;\xi_2)\,,\;&&\; \xi_{1,2}\;\epsilon\;
      \Re:\nonumber\\
      \psi(\xi;\alpha) &=& e^{i\alpha} |z \rangle \nonumber\\ 
      &=&\exp\left(i\,\alpha + i\left(\xi_2 \hat{q} -
       \xi_1\hat{p}\right)\right) | 0\rangle \nonumber\\
       &=& \exp\left(i\,\alpha -\frac{i}{2} \xi_1 \xi_2\right)
       \exp(i\;\xi_2\hat{q}) \exp\left(-i\;\xi_1 \hat{p}\right)
        | 0 \rangle \nonumber\\
       &=&\exp \left(i\;\alpha + 
\frac{i}{2} \xi_1\xi_2\right)\exp\left(-i\;\xi_1\hat{p}\right)
        \exp(i\;\xi_2\hat{q})\, |0 \rangle \,.
       \end{eqnarray}

\noindent
(Note that, as in eqn.(4.2), $\psi(\xi;\alpha)$ is a vector in ${\cal H}$ parametrised 
by $\xi$ and $\alpha$, not a wavefunction).  These various equivalent forms facilitate 
further calculations.

The expectation values of $\hat{q}$ and $\hat{p}$  in these states are
     \begin{eqnarray}
     (\psi(\xi;\alpha), \hat{q}\;\psi(\xi;\alpha)) &=& \xi_1,\nonumber\\
      (\psi(\xi;\alpha), \hat{p} \;\psi(\xi;\alpha)) &=& \xi_2\,.
      \end{eqnarray}

Now we compute the vectors $u_{\mu}(\xi;\alpha)$ and their projections 
$u^{\perp}_{\mu}(\xi;\alpha)$ orthogonal to $\psi(\xi;\alpha)$, as defined 
in eqn.(4.5):
     \begin{eqnarray}
     u_1 &=& \partial_1 \psi = -i\left(\hat{p} -\frac{1}{2} \xi_2\right)  
          \psi\;,\nonumber\\
      u_2 &=& \partial_2 \psi = i \left(\hat{q} -\frac{1}{2} \xi_1\right)
     \psi;\nonumber\\
     u^{\perp}_1 &=& -i \left(\hat{p} - \xi_2\right)\psi\,,\;\;
      \;\;u^{\perp}_2 = i\left(\hat{q} -\xi_1\right)\psi\,.
      \end{eqnarray}

\noindent
Here we used eqn.(5.5), and for simplicity omitted the arguments 
$\xi,\alpha$ in $\psi, u_{\mu}, u^{\perp}_{\mu}$.  The inner products 
among the $u^{\perp}_{\mu}$ involve the fluctuations in $\hat{q}$ and 
$\hat{p}$ and the cross term.  After easy calculations we find:
     \begin{eqnarray}
     \left(u^{\perp}_1, u^{\perp}_1\right) &=&
     \left(\psi, \left(\hat{p} -\xi_2\right)^2\psi\right) =
     (\Delta p)^2 = 1/2\;,\nonumber\\
      \left(u_1^{\perp}, u^{\perp}_2\right) &=& -\left(\psi, \left(\hat{p}
      -\xi_2\right) \left(\hat{q} -\xi_1\right) \psi\right) =       i/2\;,\nonumber\\
       \left(u_2^{\perp}, u_2^{\perp}\right) &=& \left(\psi, \left(\hat{q}
       -\xi_1\right)^2\psi\right) = (\Delta q)^2 = 1/2 \,.
       \end{eqnarray}

\noindent
Therefore the induced metric tensor in the $\xi_1-\xi_2$ plane, defined 
in eqn.(4.7), is
     \begin{eqnarray}
      g_{\mu\nu}(\xi) = \frac{1}{2} \delta_{\mu\nu}\;,
      \end{eqnarray}

\noindent
namely it is the ordinary Euclidean metric on $\Re^2$.  Constrained geodesics in 
this case are just determined by straight lines in the $\xi$-plane, as all $\Gamma$'s 
vanish:
     \begin{eqnarray}
     z(s) &=& z_0 + z_1 s\,,\;\;\;z_{0,1} = \frac{1}{\sqrt{2}}(q_{0,1} 
     + i\;p_{0,1}):\nonumber\\
     \xi_1(s) &=& q_0 + q_1 s\,,\;\;\;\xi_2(s) = p_0 + p_1 s\,.
     \end{eqnarray}

At the Hilbert space level, a constrained geodesic ${\cal C}_{\rm constr.geo.}$ 
can be taken to be a curve within the family of coherent states
     \begin{eqnarray}
     {\cal C}_{\rm constr.geo.} = \{\Psi(s) = |z_0 +z_1 s\rangle \}\,.
     \end{eqnarray}

\noindent
(Here we have omitted an $s$-dependent phase $\alpha(s)$).  Each vector $\Psi(s)$ 
along this curve is a (pure) coherent state, and can not be written as a linear 
combination of two fixed states as in eq.(2.6); so it is immediately clear that 
this is not a free geodesic at all.

Now we examine whether this constrained geodesic is a null phase curve.  We find, 
using the criterion (3.4):
     \begin{eqnarray}
     \arg (\Psi(s), \Psi(s^{\prime})) &=&\arg \langle z_0+z_1 s |z_0 + z_1           
s^{\prime}\rangle  \nonumber\\
     &=&\arg \left(\exp\left\{\left(z^{*}_0 + z_1^{*} s\right)
     \left(z_0+z_1 s^{\prime}\right)\right\}\right)\nonumber\\
     &=&\arg \left(\exp\left(z^{*}_0 z_1 s^{\prime} + z_0 z_1^{*}           s\right)\right)\nonumber\\
     &=& (s^{\prime} - s) \mbox{Im} z^{*}_0 z_1  \,.
      \end{eqnarray}

\noindent
This is a separable function of $s^{\prime}$ and $s$, so we do have a null phase 
curve.  We can go from the above ${\cal C}_{\rm constr.geo.}$ to a horizontal 
curve by adding a phase:
     \begin{eqnarray}
     {\cal C}^{(h)}_{\rm constr.geo.} = \left\{\Psi^{\prime}(s) =
      \exp \left(- i\,s\,\mbox{Im}\,z_0^{*} z_1\right) \Psi(s)\right\}\,,
      \end{eqnarray}

\noindent
and then we find that {\em any two} points on this curve 
are in phase, as expected.

The generalised connection (3.7) in this example now states: if 
$|z_1\rangle , |z_2\rangle , \ldots, |z_n\rangle $ are any $n$ pure coherent states given 
by choosing $n$ points in the complex plane, and we join these points 
successively by straight lines in the complex plane,
 so that all along in Hilbert space we deal with individual coherent states 
and never with superpositions of them, we have:
     \begin{eqnarray}
     \varphi_g \left[\begin{array}{c}n\mbox{-sided plane polygon with vertices}\\
     \mbox{at the coherent states}\;z_1,z_2,\ldots,z_n\end{array}\right] =
      -\arg \Delta_n (|z_1\rangle , |z_2\rangle , \ldots, |z_n\rangle ) \,.
     \end{eqnarray}

\noindent
The case $n=3$ leads to the area formula for the geometric phase for a triangle 
in the plane, a very familiar result\cite{scvs}.
From our point of view, the present example is a significant generalisation 
of the original connection (2.9).

Going further, it is easy to convince oneself that in this example the most 
general null phase curve arises in the above manner; in other words, a given 
one-parameter family of coherent states $\{|z(s)\rangle\}$ obeys the separability 
condition (3.4) if and only if $\mbox{Im}\,z(s)$ is a linear inhomogeneous 
expression in $\mbox{Re} z(s)$, so $z(s)$ describes a straight line in the complex 
plane as $s$ varies.

\noindent
(c) {\bf Centred Gaussian pure states}

This example again deals with one canonical pair $\hat{q},\hat{p}$.  It is now 
more convenient to work with wavefunctions in the Schrodinger representation, and 
not with abstract ket  vectors.  The submanifold ${\cal M}\subset {\cal H}$ consists 
of normalised Gaussian wavefunctions parametrised by two real variables 
$\xi_1,\xi_2$ and a phase angle $\alpha$ defined as follows:
     \begin{eqnarray}
     \psi(\xi;\alpha;q) &=& \left(\frac{\xi_2}{\pi}\right)^{1/4}
     \exp\left\{i \alpha + \frac{i}{2} (\xi_1 +
           i\;\xi_2)q^2\right\}\;,\nonumber\\
     &&\xi_1 \in (-\infty, \infty)\,,\;\; \;\xi_2 \in(0,\infty)\,,\;\;\;\alpha
     \in [0,2 \pi)\,.
      \end{eqnarray}

\noindent
Normalisability requires that $\xi_2$ be strictly positive, so the combination 
$\xi_1+i\;\xi_2$ is a variable point in the upper half complex plane.  The wave 
functions $u_{\mu}(\xi;\alpha;q)$ are:
     \begin{eqnarray} 
     u_1(\xi;\alpha;q) = \frac{\partial}{\partial \xi_1} \psi(\xi;\alpha;q) 
      &=&\frac{i}{2}\, q^2\, \psi(\xi;\alpha;q),\nonumber\\
     u_2(\xi;\alpha;q) = \frac{\partial}{\partial \xi_2} \psi(\xi;\alpha;q)  
     &=& \frac{1}{2}\left(-q^2 + \frac{1}{2\xi_2}\right) \psi(\xi;\alpha;q)\,.
     \end{eqnarray}

\noindent
It is clear that to obtain the components $u^{\perp}_{\mu}$ of $u_{\mu}$ orthogonal 
to $\psi$, and later to compute the inner products 
$\left(u^{\perp}_{\mu},u^{\perp}_{\nu}\right)$, we need the expectation values 
of $q^2$ and $q^4$ in the state $\psi$.  
These are (omitting for simplicity the arguments of $\psi$):
     \begin{eqnarray}
     (\psi,q^2 \psi) &=& \left(\frac{\xi_2}{\pi}\right)^{1/2}
      \!\!\int\limits^{\infty}_{-\infty} \!dq\, q^2\, e^{-\xi_2 q^2}=
      \frac{1}{2\xi_2}\,,\nonumber\\
      (\psi,q^4 \psi)&=& \left(\frac{\xi_2}{\pi}\right)^{1/2}
      \!\!\int\limits^{\infty}_{-\infty}\! dq\, q^4\, e^{-\xi_2 q^2} =
      \frac{3}{4 \xi^2_2}\,.
      \end{eqnarray}

\noindent
Now the necessary inner products and projections are easily found:
     \begin{eqnarray}
     (\psi, u_1) &=& \frac{i}{4\xi_2}\,,\;\;\;(\psi, u_2) = 0\,;\nonumber\\
      u^{\perp}_1 &=& \frac{i}{2} \left(q^2 - \frac{1}{2\xi_2}\right)\psi\, ,
      \nonumber\\
      u^{\perp}_2 &=& u_2 = -\frac{1}{2}\left(q^2 -\frac{1}{2\xi_2}\right) 
      \psi\, ;\nonumber\\    
      \left(u^{\perp}_1, u^{\perp}_1\right) &=& \frac{1}{4}
      \left(\psi, \left(q^2 -\frac{1}{2\xi_2}\right)^2\! \psi\right)
       = \frac{1}{8\xi^2_2}\, ,\nonumber\\
       \left(u^{\perp}_1, u^{\perp}_2\right) &=&\frac{i}{4}
       \left(\psi, \left(q^2 -\frac{1}{2\;\xi_2}\right)^2\!\psi\right) =
          \frac{i}{8\xi^2_2}\, ,\nonumber\\
       \left(u^{\perp}_2, u^{\perp}_2\right) &=&
        (u_2, u_2) = \frac{1}{4}\left(\psi, \left(q^2 -\frac{1}{2\;\xi_2}
        \right)^2\!\psi \right) = \frac{1}{8\;\xi^2_2}\, .
       \end{eqnarray}

\noindent
From these results we obtain the induced metric over 
$M=\pi[{\cal M}]\subset R$, described in the upper half 
complex plane by the metric tensor
     \begin{eqnarray}
     g_{\mu\nu}(\xi) = \mbox{Re}\left(u^{\perp}_{\mu}(\xi;\alpha),
     u^{\perp}_{\nu}(\xi;\alpha)\right) =  \frac{1}{8\xi^2_2}\,\delta_{\mu\nu}\, .
     \end{eqnarray}

\noindent
This is the well-known form of the Lobachevskian metric in this model of 
Lobachevsky space\cite{loba}.  Dropping the numerical factor 1/8 for simplicity, the 
line element in the upper half plane is given by
     \begin{eqnarray}
     ds^2  = \frac{1}{\xi^2_2}\,\left(d\xi_1^2 + d\xi^2_2\right)\,,
     \end{eqnarray}

\noindent
and we must find the corresponding geodesics.

First we compute the nonvanishing $\Gamma$'s.  The inverse of 
$(g_{\mu\nu}(\xi))$ has components
     \begin{eqnarray}
     g^{11}(\xi) = g^{22}(\xi) = \xi^2_2\,,\;\;\; g^{12}(\xi) = 0\,.
     \end{eqnarray}

\noindent
We easily find that the nonvanishing $\Gamma$'s are
     \begin{eqnarray}
     \Gamma^1\,_{12}(\xi) = \Gamma^2\,_{22}(\xi) = -\Gamma^2\,_{11}(\xi)
     = - \, \frac{1}{\xi_2}\,.
     \end{eqnarray}

\noindent
Using these in eqn.(4.8) we find the following ordinary differential 
equations to determine geodesics:
    \begin{mathletters}
     \begin{eqnarray}
     \ddot{\xi}_1 - \frac{2}{\xi_2}\,\dot{\xi}_1 \dot{\xi}_2  &=& 0 \,,\\
     \ddot{\xi}_2 + \frac{1}{\xi_2}\,\left(\dot{\xi}^2_1 - \dot{\xi}^2_2\right) 
     &=& 0\,.
     \end{eqnarray}
     \end{mathletters}

We can exploit the fact that these differential equations lead to the consequence
     \begin{eqnarray}
     \frac{1}{\xi_2^2} \!\left(\dot{\xi}^2_1 + \dot{\xi}^2_2\right)  
=\mbox{constant}\,,     
     \end{eqnarray}

\noindent
the value of the constant depending on the particular geodesic.  After elementary 
analysis we find that there are two families of geodesics:
     \begin{mathletters}
     \begin{eqnarray}
     \mbox{Type I:}\;\;\;\; && \xi_1 =\mbox{constant}\, ,\;\;\; \xi_2 =a e^{bs} 
\,,\nonumber\\
     &&a>0 \,,\;\;\; s\,\in \Re\,;\\
     \mbox{Type II:}\;\;\;\; && \xi_1 = c + R \cos\! f(s) \,,\;\;\; \xi_2 = R \sin\! 
f(s)\, ,
     \nonumber\\
     &&f(s) = 2 \tan^{-1} (a e^{bs}) \,,\nonumber\\
     &&c\in \Re \,,\;\;\; R>0\,,\;\;\; a>0\,,\;\;\; b>0\,,\;\;\; s\in \Re\,.
     \end{eqnarray}
     \end{mathletters}

\noindent
These are both in affinely parametrised form.  In Type II it is simpler to 
pass to a nonaffine angle type parameter $s\in (0,\pi)$, and replace eqn.(5.24b) by:
     \begin{eqnarray}
     \mbox{Type II:}\;\;\;\;&& \xi_1 = c + R \cos s\,,\;\;\; \xi_2 = R\sin s 
      \,,\nonumber\\
      &&c\in\Re\,,\;\;\; R>0 \,,\;\; \;0<s<\pi\,.
      \end{eqnarray}

\noindent
Type I geodesics are straight semi infinite lines parallel to the $\xi_2$ axis.  
Type II geodesics are semicircles centered on the $\xi_1$ axis and lying above this 
axis.

In each case we can now ask whether a constrained geodesic in $M$ is a null phase 
curve.  As in the previous example of coherent states, here too we emphasize that 
we are concerned with curves within the manifold of centred normalised Gaussian 
wavefunctions, and at no stage with linear combinations of such wavefunctions. 
We look at the two types of constrained geodesics in turn and find these results 
(after simple reparametrisations):
     \begin{mathletters}
     \begin{eqnarray}
     \mbox{Type I:}\;\;\;\; && \Psi(s) = \psi(\xi_1 =a,\, \xi_2=bs;\, 
\alpha(s)):\nonumber\\
      && \;\;\;\;\;\;\arg (\Psi(s), \Psi(s^{\prime})) = 0\, ;\\
      \mbox{Type II:}\;\;\;\;&& \Psi(s) = \psi(\xi_1 =c+R \cos s,\, \xi_2=R \sin s;
     \, \alpha(s)) :\nonumber\\
      &&\arg(\Psi(s), \Psi(s^{\prime})) = \frac{1}{4}(s - s^{\prime})\,.
      \end{eqnarray}
      \end{mathletters}

\noindent
(In both cases the choice of phase angle $\alpha(s)$ is irrelevant).  So in both 
cases the criterion (3.4) is obeyed; and both types of curves in $M$ arising from 
the two types of geodesics in the upper half $\xi$ plane are simultaneously 
constrained geodesics and null phase curves.

The statement of the generalised connection (3.7) is clear, and for illustration 
we consider the case of just three vertices.  
Let $A, A^{\prime}$ and $A^{\prime\prime}$ be any three points in the upper half 
complex plane; and for any choices of phases $\alpha$ consider the three normalised 
centered Gaussian states $\psi(A;\alpha), \psi(A^{\prime}; \alpha^{\prime})$ 
and $\psi(A^{\prime\prime};\alpha^{\prime\prime})$.  
Join $A$ to $A^{\prime}, A^{\prime}$ to $A^{\prime\prime}$ and 
$A^{\prime\prime}$ to $A$ by a geodesic of Type I or Type II 
as appropriate in each case.  This can always be done, and we obtain a hyperbolic 
triangle.  In $M$ we obtain a `triangle' with vertices $\rho(A)=\pi(\psi(A;\alpha))$ 
etc., and whose sides are constrained geodesics; and we can state:
     \begin{eqnarray}
     \varphi_g\left[
\begin{array}{c}
\mbox{`triangle' in}\; M \;\mbox{with vertices}\\
\;\rho(A), \rho(A^{\prime}), \rho(A^{\prime\prime})\\
      \mbox{and sides as constrained geodesics}
\end{array}\right]
 = -\arg \Delta_3
      (\psi(A;\alpha), \psi(A^{\prime};\alpha^{\prime}),       
         \psi(A^{\prime\prime};\alpha^{\prime\prime}))\,.
      \end{eqnarray}

\noindent
An application of this result has been used elsewhere\cite{gouyprl} to show that the 
classical Gouy phase\cite{gouy}
 in wave optics is related to a Bargmann invariant and hence is a geometric 
phase.

\noindent
(d)  {\bf A subset of two-mode coherent states}

   In the previous two examples we found that while constrained geodesics differed 
from free geodesics, they were nevertheless null phase curves and so led to important 
instances of eqn.(3.7).  This is however fortuitous; the really important objects 
for our purposes are the null phase curves, and in a given situation constrained 
geodesics may well not be such curves.  In our fourth and final example, dealing 
with a subset of states for a two-mode system, we will find that this is just what 
happens.  However we will be able to completely determine all null phase curves 
directly,  so that the generalisation (3.7) can be meaningfully stated.

For a two-mode system with creation and annihilation operators $\hat{a}_j^+, 
\hat{a}_j$
  obeying the standard commutation relations
     \begin{eqnarray}
     \left\protect[\hat{a}_j, \hat{a}_{k}^{\dag} \right\protect]=\delta_{jk}\,,\;\;\; 
[\hat{a}_j, \hat{a}_k] = \left\protect[\hat{a}^{\dag}_j, 
a^{\dag}_k\right\protect]=0,\;\;\;
     j,k=1,2
     \end{eqnarray}

\noindent
the general coherent state is labelled by two independent complex numbers 
arranged as a column vector     $\mbox{\boldmath $z$} = (z_1, z_2)^T :$
      \begin{eqnarray}
     |\mbox{\boldmath $z$} \rangle  &=& \exp \left(-\frac{1}{2}   
   \mbox{\boldmath $z$} ^{\dag}\mbox{\boldmath $z$} + 
       z_1 \hat{a}^{\dag}_1 + z_2  
                    \hat{a}^{\dag}_2\right) |0 \rangle  ,\nonumber\\
      \hat{a}_j |\mbox{\boldmath $z$} \rangle 
     &=& z_j |\mbox{\boldmath $z$} \rangle \,,\;\;\; j=1,2\,.    
      \end{eqnarray}

\noindent
Within this family of all normalised coherent states we now define a submanifold 
(of real dimension three including an overall phase), an ``$S^2$-worth of states'', 
by taking $\theta, \phi$ to be spherical polar angles on a sphere $S^2$ and setting 
$z_1$ and $z_2$ equal to the following:
     \begin{eqnarray}
     z_1 = \cos \theta\,,\;\;\; z_2 = e^{i\phi} \sin \theta\,,\;\;\; 0\leq \theta 
    \leq \pi\,,\;\;\; 0\leq \phi \leq 2\pi\,.
     \end{eqnarray}

\noindent
Therefore the submanifold ${\cal M}\subset {\cal H}$ is parametrised by 
$\theta, \phi$ and  a phase $\alpha$ and we write:
     \begin{eqnarray}
     {\cal M}=\left\{\psi(\theta,\phi;\alpha) = e^{i\alpha} |\cos \theta,
      e^{i\phi} \sin \theta\rangle \;\;\big|\;\; 0\leq \theta\leq \pi\,,\;\;\;
      0\leq\phi,\, \alpha\leq 2\pi\right\} \subset {\cal H}\, ,\nonumber\\
      \end{eqnarray}
\noindent
where the ket on the right is a particular two-mode coherent state with 
$\mbox{\boldmath $z$} ^{\dag} \mbox{\boldmath $z$} =1$
    \begin{eqnarray}
    \psi(\theta,\phi;\alpha) = \exp\left(i\alpha + \hat{a}_1^{\dag}
    \cos \theta + \hat{a}^{\dag}_2 e^{i \phi} \sin \theta -1/2\right) |0\rangle \,.
    \end{eqnarray}

\noindent
Omitting the arguments $\theta, \phi, \alpha$ for simplicity, we easily find:
     \begin{mathletters}
     \begin{eqnarray}
     u_{\theta} = \frac{\partial}{\partial \theta} \psi &=&
      \left(-\sin \theta \;\hat{a}^{\dag}_1 + e^{i\phi}\cos \theta\;
      \hat{a}_2^{\dag}\right)\psi \,,\nonumber\\
      u_{\phi} = \frac{\partial}{\partial \phi} \psi &=&
      i\,e^{i \phi} \sin \theta \,\hat{a}^{\dag}_2\, \psi \,;\\
      (\psi, u_{\theta}) = 0\,,\;\;\; && (\psi, u_{\phi}) = i\, \sin^2 \theta 
             \,;\\
       u^{\perp}_{\theta} = u_{\theta}\,,\;\;\; u^{\perp}_{\phi}&=&
       i\, \sin \theta \left(e^{i \phi}\hat{a}^{\dag}_2 -
       \sin \theta \right) \psi\,.
       \end{eqnarray}
       \end{mathletters}

\noindent
Repeatedly exploiting the eigenvector relation (5.29) and its adjoint, 
we compute the inner products among the vectors  in  eq.(5.33c):
     \begin{eqnarray}
     \left(u^{\perp}_{\theta}, u^{\perp}_{\theta}\right) = 1\,,\;\;\; 
\left(u^{\perp}_{\theta},
     u^{\perp}_{\phi}\right) = i\, \cos \theta  \sin \theta \,,\;\;\;
      \left(u^{\perp}_{\phi}, u^{\perp}_{\phi}\right) = \sin^2 \theta\,.
      \end{eqnarray}

\noindent
Taking the real parts here, we see that the metric induced on 
$M=\pi[{\cal M}]\sim S^2$ in ${\cal R}$, parametrised by angles 
$\theta$ and $\phi$, is just the usual rotationally invariant one:
     \begin{eqnarray}
     g_{\theta\theta}(\theta, \phi) = 1 \,,\; \;\;g_{\theta\phi} = 0\, ,\;\;\;
     g_{\phi\phi}(\theta, \phi) = \sin^2 \theta\,.
     \end{eqnarray}

\noindent
The corresponding constrained geodesics are therefore simply great circle arcs.  
The question is whether they lead to null phase curves in $M$ and ${\cal M}$.

A general parametrised great circle arc on $S^2$ is traced out by an $s$-dependent 
unit vector $\hat{n}(s)$ with polar angles $\theta(s), \phi(s)$:
     \begin{eqnarray}
     \hat{n}(s) = \hat{a} \cos  s+ \hat{b} \sin s &=&
     \left(\sin \theta(s) \cos \phi(s), \;\sin \theta(s) \sin \phi(s),\;
      \cos \theta(s)\right) \,,\nonumber\\
     && \hat{a}, \hat{b} \in S^2\,,\;\;\; \hat{a} \cdot \hat{b} =0\,.
      \end{eqnarray}

\noindent
The corresponding constrained geodesic 
${\cal C}_{{\rm constr. geo.}}\subset {\cal M}$ (omitting the phase $\alpha$) 
is the curve of coherent states
     \begin{eqnarray}
     \Psi(s) &=& |z_1 (s), z_2(s) \rangle \,,\nonumber\\
     z_1 (s) = \cos \theta(s) &=& a_3 \cos s + b_3 \sin s\,,\nonumber\\
     z_2(s) = e^{i\, \phi(s)}\sin \theta(s)&=&
     (a_1 +i\, a_2) \cos s + (b_1 + i\, b_2)\sin s\,.
      \end{eqnarray}

\noindent
To see whether this is a null phase curve we compute the phase of 
$(\Psi(s), \Psi(s^{\prime}))$:
     \begin{eqnarray}
     \arg\left(\Psi(s), \Psi(s^{\prime})\right) &=&
     \arg \langle z_1 (s), z_2(s)| z_1(s^{\prime}), z_2(s^{\prime}) \rangle 
   \nonumber\\
      &=&\arg \left\{\exp\left(z_1(s) z_1(s^{\prime}) + z_2(s)^* 
                  z_2(s^{\prime})\right)\right\}\nonumber\\
     &=&\arg\left\{\exp\left([(a_1-i\, a_2)\cos s +(b_1 -i\, b_2)\sin s]
         \right.\right.\nonumber\\
     &&\left.\left.\left\protect[(a_1+i\, a_2)\cos s^{\prime} + 
        (b_1 + i\, b_2)\sin s^{\prime}\right\protect]\right)\right\}\nonumber\\
     &=& (\hat{a}_{\land} \hat{b})_3 \, \sin(s^{\prime} - s)\,.
     \end{eqnarray}

\noindent
Unless it vanishes, this is not a separable function of $s^{\prime}$ and $s$.  
We conclude that the geodesic (5.36) on $S^2$ leads to a constrained geodesic 
${\cal C}_{\rm constr.geo.}\subset {\cal M}$ which is in general not a null 
phase curve.  The only exception is when $(\hat{a}_{\land}\hat{b})_3=0$, that is, 
the geodesic (5.36) on $S^2$ lies on a meridian of longitude, with 
$\hat{a}_{\land}\hat{b}$ being a vector in the 1-2 plane.

On the other hand, in this example it is quite easy to explicitly find all null phase 
curves on $M$(and ${\cal M})$!  Let $\Gamma=\{\hat{n}(s)\}\subset S^2$ be given, and 
let us consider the induced curve ${\cal C}_{\Gamma}$ in ${\cal M}$:
     \begin{eqnarray}
      {\cal C}_{\Gamma} &=& \{\Psi_{\Gamma}(s) = |n_3(s), n_1(s) + 
      i\,n_2(s) \rangle  =\nonumber\\
      && \exp (- \frac{\scriptsize 1}{\scriptsize 2} + n_3(s)\; 
\hat{a}_1^{\dag} + (n_1(s)
       + i\,n_2(s))\hat{a}_2^{\dag}) |0 \rangle  \}\,.
       \end{eqnarray}

\noindent
We find that
     \begin{eqnarray}
     \arg\left(\Psi_{\Gamma}(s), \Psi_{\Gamma}(s^{\prime})\right) =
     \left(\hat{n}(s)_{\land} \hat{n}(s^{\prime})\right)_3\,.
     \end{eqnarray}

\noindent
This will be a separable function of $s^{\prime}$ and $s$ if and only if, for some 
constants $\beta$ and $\gamma$, we have
      \begin{eqnarray}
       n_2(s) = \beta\;n_1(s) + \gamma\,.
       \end{eqnarray}   

\noindent
The geometrical interpretation of this is that the projection of $\Gamma$ on 
the 1-2 plane must be a straight line.  In that case ${\cal C}_{\Gamma}$ is 
indeed a null phase curve in ${\cal M}$, as we have
     \begin{eqnarray}
     \arg (\Psi_{\Gamma}(s), \Psi_{\Gamma}(s^{\prime})) = \gamma(n_1(s) - 
n_1(s^{\prime})) \,,
     \end{eqnarray}

\noindent
which is separable in $s^{\prime}$ and $s$.  One can easily see that each such 
$\Gamma$ is a latitude circle arc on $S^2$ corresponding to (i.e., perpendicular to) 
some axis lying in the 1-2 plane; and given any two points on $S^2$, we can always 
connect them by such a $\Gamma$.  In other words, such $\Gamma$ are intersections 
of $S^2$ with planes perpendicular to the 1-2 plane.  When such a latitude circle 
arc is also a great circle arc, we recover the result of the previous paragraph.

The upshot of this example is that here we have a nontrivial illustration of the 
difference between constrained geodesics and null phase curves.  However, since we 
have been able to find all of the latter, and any two points in ${\cal M}$ can be 
connected by some null phase curve, we have succeeded in providing a nontrivial 
two-mode example of the generalised connection (3.7), but not using constrained 
geodesics.
\section{RAY SPACE AND DIFFERENTIAL GEOMETRIC FORMULATIONS}
Very soon after the discovery of the geometric phase, the differential geometric
expressions of its structure and significance were brought 
out\cite{simon,anandan,samuel,nonabelian,page}, by relating it to anholonomy and 
curvature in
a suitable  Hermitian line  bundle on quantum mechanical ray space. In this
section we provide a brief discussion of the properties and uses of the new concept
of  null phase curves  at ray space level and also in the 
differential geometric language. Only necessary background material will be 
recalled, and derivations
omitted. Since they may be useful for practical calculations, where possible local
coordinate expressions of important differential geometric objects will be given.

From the preceding sections it is evident that for our purposes it is important to
deal with open  null phase curves in general, since it is through them that 
the connection (3.7) of the Bargmann invariants to geometric phases is made. Their
definition (3.4) in terms of Hilbert space lifts is quite simple. Nevertheless it is
of interest to develop a direct ray space formulation; this can be done essentially
via the Bargmann invariants themselves. From their definition (2.8), it is clear
that any $\bigtriangleup_2$ is real nonnegative, while  $\bigtriangleup_n$'s for 
$n \ge 3$ are
in general complex. On the other hand it is also known that any $\bigtriangleup_n$ for $n
\ge 4$ can be written as the ratio of a suitable product of $\bigtriangleup_3$'s and a
suitable product of $\bigtriangleup_2$'s:
\begin{equation}
\bigtriangleup_n(\psi_1,\psi_2,\cdots,\psi_n) = \prod^n_{j=3}
\bigtriangleup_3(\psi_1,\psi_{j-1},\psi_j)/\prod^n_{j=4}
\bigtriangleup_2(\psi_1,\psi_{j-1}).
\end{equation}
In this sense  the three-vertex Bargmann 
invariant  
 $\bigtriangleup_3$  is the basic or primitive one as far as phases are concerned.  
(The basic cyclic invariance of $\bigtriangleup_n(\psi_1,\psi_2,\cdots,\psi_n)$
is not manifest in eq.(6.1), but it is not lost either).
Guided by 
these facts we give now a direct ray space characterisation of null phase curves.

If $C=\{\rho(s)\} \subset {\cal R}$ is a null phase curve and ${\cal C}^{(h)} =
\{\psi^{(h)}(s)\}$ is a horizontal Hilbert space lift obeying eq.(3.4), we
see immediately that for any choices of parameter values $s,s',s''$,
\begin{eqnarray}
\bigtriangleup_3 (\psi^{(h)}(s), \psi^{(h)}(s'), \psi^{(h)}(s'')) \ = \ \mbox{Tr}
\{\rho(s)\rho(s') \rho(s'')\} \ = \ \mbox{real and}\; \ge 0\, ;  
\end{eqnarray}
and so also for any $n$  parameter values $s_1, s_2, \ldots , s_n$,
from eqn.(6.1),
\begin{eqnarray}
\bigtriangleup_n (\psi^{(h)}(s_1), \ldots , \psi^{(h)} (s_n)) = \mbox{Tr} \{\rho(s_1)
\ldots \rho(s_n)\} \ = \ \mbox{real and} \;\ge 0\,. 
\end{eqnarray}

\noindent
As a consequence, by differentiation with respect to $s_2, \ldots , s_n$ we have
:
\begin{eqnarray}
\mbox{Tr}\left\{\rho(s_1) \frac{d\rho(s_2)}{ds_2} \ldots \frac{d\rho(s_n)}{ds_n} 
\right\} \ = \ {\rm real}\,.
\end{eqnarray}

\noindent
Now it is known that the geometric phase for any connected portion of any $C$
can be expressed directly in terms of $\rho(s)$ as follows, whether or not $C$
is a null-phase curve:
\begin{eqnarray}
\varphi_g [\rho(s_1)\; {\rm to } \;\rho(s_2)\; {\rm along}\; \ C] \ = \ {\rm arg}
\left[\mbox{Tr}\left\{\rho(s_1) P \left({\rm exp} \int^{s_2}_{s_1}\! ds 
\frac{d\rho(s)}{ds}\right) \right\}\right] \nonumber \\
= {\rm arg} \left[1+ \sum^\infty_{n=1} \int^{s_2}_{s_1} ds'_n \int^{s'_n}_{s_1}
ds'_{n-1} \ldots \int^{s'_2}_{s_1} ds'_1\mbox{Tr} \left\{\rho(s_1) 
\frac{d\rho(s'_n)}{ds'_n} \ldots \frac{d\rho(s'_1)}{ds'_1} \right\} \right]\,, 
\end{eqnarray}
where $P$ is the  ordering symbol placing later parameter values to the
left of earlier ones. If eq.(6.2) holds on $C$ (and so as a consequence
eqs.(6.3,4) as well), we see that at every stage only real quantities are
involved, the geometric phase in eq.(6.5) vanishes, and $C$ is a null phase
curve. This leads to the ray space characterisation of null phase curves we are 
seeking : 
\begin{eqnarray}
C \ = \ \{\rho(s)\} \subset {\cal R}\;\; {\rm is\ a\ null\ phase\ curve}\;\;
&\Longleftrightarrow& \nonumber \\
{\rm Tr}\{\rho(s) \rho(s') \rho(s'')\} \ &=& \ {\rm\ real\ nonnegative,\ any} \ s,s',
s''\,.
\end{eqnarray}
Turning now to the specific differential geometric aspects, it is well known
that while the dynamical phase $\varphi_{\rm dyn}[{\cal C}]$ is an additive
quantity, $\varphi_g[C]$ does not have this property. On the manifold of unit
vectors in Hilbert space ${\cal H}$, there is a one form $A$ such that
\begin{eqnarray}
\varphi_{\rm dyn} [{\cal C}] \ = \ \int_{\cal C} A\,. 
\end{eqnarray}
However, referring to the projection $\pi: {\cal H} \rightarrow {\cal R},\; A$ is
{\it not} the pull-back via $\pi^{\,*}$ of any one-form on the space of unit rays; and
$\varphi_g[C]$ is {\it not} the integral along $C$ of any one-form on ${\cal
R}$. In fact this lack of additivity can be expressed via the Bargmann invariant
$\bigtriangleup_3$. If $C_{12}$ connects $\rho_1$ to $\rho_2$ in ${\cal R}$ and $C_{23}$
connects $\rho_2$ to $\rho_3$, than $C_{12} \cup C_{23}$ runs from $\rho_1$ to $\rho_3$
and
\begin{eqnarray}
\varphi_g[C_{12} \cup C_{23}] \ &=& 
\ \varphi_g[C_{12}] + \varphi_g[C_{23}] - B_3(\psi_1,\psi_2,\psi_3),\nonumber \\
 B_3(\psi_1,\psi_2,\psi_3)&=&
 {\rm arg}\bigtriangleup_3(\psi_1,\psi_2,\psi_3).
\end{eqnarray}
More generally, for an (generally) open curve consisting of $(n-1)$ pieces 
$C_{12},C_{23}, \cdots C_{n-1,n}\/$ joining $\rho_1\/$ to $\rho_2\/$,
 $\rho_2\/$ to $\rho_3\/$, $\cdots , \rho_{n-1}\/$ to $\rho_n\/$, we generalize 
 eq. (6.8) to the following:
 \begin{eqnarray}
\varphi_g[C_{12}\cup C_{23} \cup \cdots \cup C_{n-1,n}]&=&\sum^{n-1}_{j=1}
\varphi_g[C_{j,j+1}]-B_n(\psi_1,\psi_2,\cdots,\psi_n),\nonumber \\
B_n(\psi_1,\psi_2,\cdots,\psi_n)&=& {\rm arg} \bigtriangleup_n
(\psi_1,\psi_2,\cdots,\psi_n)\nonumber \\
&=&\sum^n_{j=3}B_3(\psi_1,\psi_{j-1},\psi_j).
 \end{eqnarray}
If we connect $\rho_n\/$ back to $\rho_1\/$ via $C_{n,1}\/$ to get a closed curve of
$n\/$ pieces, then we have the specific result:
\begin{equation}
\varphi_g[C_{12}\cup C_{23} \cup \cdots \cup C_{n-1,n}, \cup C_{n,1}]=\varphi_g[C_{12}]+
\varphi_g[C_{23}]+\cdots +\varphi_g[C_{n,1}] - B_n(\psi_1,\psi_2,\cdots,\psi_n).
\end{equation}
Compared to eq.~(6.9), we have one extra $\varphi_g\/$ term on the right but the Bargmann
phase term $B_n\/$ is the same. We see that the lack of additivity shown in all 
eqs.(6.8,6.9,6.10) is due to the Bargmann pieces.
There is however an exception to this general nonadditivity, which occurs in~(6.8) when
$\rho_3 = \rho_1$ and $C_{12} \cup C_{23}$ is a closed loop. Then we find :
\begin{eqnarray}
\partial (C_{12} \cup C_{23})  &=&  0, \quad \rho_3=\rho_1 :\nonumber \\
\varphi_g[C_{12} \cup C_{21}] & = & \varphi_g[C_{12}] + \varphi_g[C_{21}]\,,
\nonumber \\
\mbox{i.e.,}\;\;\;\;  \varphi_g [C_{12}] & = & \varphi_g [C_{12} \cup C_{21}] 
- \varphi_g [C_{23}]\,.
\end{eqnarray}
In the past this result has been used\cite{samuel} to relate $\varphi_g[C]$ for an 
open 
$C$ to $\varphi_g[C \cup C']$ for a closed $C \cup C'$ by choosing $C'$ to be a free
geodesic, for then $\varphi_g[C'] = 0$. Now we can generalise this process: if
$C$ is an open curve from $\rho_1$ to $\rho_2$ in ${\cal R}$, and $C'$ is any
null phase curve from $\rho_2$ back to $\rho_1$, we have the result
\begin{eqnarray}
\varphi_g[{\rm open \ curve}\;\; C] \ = \ \varphi_g [{\rm closed \ loop}\;\; C 
\cup C'] \,.\end{eqnarray}
This is the most general way in which an open curve geometric phase can be
reduced to a closed loop geometric phase.
More generally, comparing eqs.~(6.9,6.10) valid for generally open and for a 
closed curve, we see that  {\it if the last piece $C_{n,1}$ is a null phase
curve} we convert an open curve geometric phase to a closed loop geometric phase:
\begin{equation}
\varphi_g[C_{12}\cup C_{23} \cup \cdots \cup C_{n-1,n}]=
\varphi_g[C_{12}\cup C_{23} \cup \cdots \cup C_{n-1,n}\cup C_{n,1}]
\end{equation}

At this point it is natural to express a closed loop geometric phase as a
suitable ``area integral'' of a two-form, both at Hilbert and ray space levels.
Whereas $A$ is  not the pull back of any one-form on ${\cal R}$, we do have $dA =
\pi^{\,*}\,\omega$, where $\omega$ is a symplectic  (closed, nondegenerate) 
two-form on ${\cal
R}$. Then, if ${\cal C}$ is a closed loop in ${\cal H}$, $\partial {\cal C} =
0$, so that $C = \pi({\cal C})$ is a closed loop in ${\cal R}$, we have
\begin{eqnarray}
\varphi_g[C] \ = \ \int_{\cal S} dA \ = \ \int_S \omega \,,
\end{eqnarray}
where ${\cal S}$ and $S=\pi({\cal S})$ are two-dimensional surfaces in ${\cal H}$ and 
${\cal R}$ respectively, with boundaries ${\cal C}$ and $ C$ :
\begin{eqnarray}
\partial {\cal S} = {\cal C} \,,\;\;\; \partial S \ = \ C\,.
\end{eqnarray}
With the help of local coordinates on ${\cal H}$ and ${\cal R}$ we get explicit
expressions for $A, dA$ and $\omega$. Around any point $\rho_0 \in {\cal R}$, and for
some chosen $\psi_0 \in \pi^{-1} (\rho_0)$, we define an (open) neighbourhood $N
\subset {\cal R}$ by
\begin{eqnarray}
N \ = \ \left\{\rho \in {\cal R}\;\;\big|\;\; \mbox{Tr}(\rho_0 \rho) > 0\right\}\,.
\end{eqnarray}
We can introduce real independent coordinates over $N$ as follows. Let
$\{\psi_0, e_1, e_2, \ldots , e_r, \ldots\}$ be an orthonormal basis for ${\cal
H}$. Then points in $N$ can be ``labelled'' in a one-to-one manner by
vectors ${\cal X} \in  {\cal H}$ orthogonal to $\psi_0$ and with norm less
than unity:
\begin{eqnarray}
\chi(\beta, \gamma) & = & \frac{1}{\sqrt{2}} \sum_r (\beta_r-i
\gamma_r)\, e_r \,, \nonumber \\
\| \chi (\beta, \gamma)\|^2 & = & \frac{1}{2} \sum_r (\beta^2_r +
\gamma_r^2) < 1 :\nonumber \\
\psi(\beta, \gamma) & = & \chi(\beta, \gamma) + \sqrt{1-\|\chi(\beta,
\gamma)\|^2\,}\, \psi_0\,, \nonumber \\
\rho \in N\; \Longleftrightarrow  \; \rho & = & \psi (\beta, \gamma)
\psi(\beta, \gamma)^\dagger\,, \;\;{\rm for\ some} \ \beta, \gamma\,. 
\end{eqnarray}
Thus the real independent $\beta$'s and $\gamma$'s, subject to
the inequality above, are local coordinates for $N$. They can be
extended to get local coordinates for $\pi^{-1} (N) \subset {\cal
H}$ by including a phase angle $\alpha$ :
\begin{eqnarray}
\psi \in \pi^{- 1} (N)\; \Longleftrightarrow\; \psi = \psi (\alpha ; \beta , \gamma)
= e^{i\alpha} \psi (\beta, \gamma) \,,\;\;\; 0 \le \alpha < 2\pi\,.
\end{eqnarray}
In these local coordinates over $N$ and $\pi^{-1}(N)$ we have the
expressions 
\begin{eqnarray}
A & = & d\alpha + \frac{1}{2} \sum_r (\gamma_r d\beta_r - \beta_r
d\gamma_r)\,, \nonumber \\
dA & = & \sum_r d\gamma_r \wedge d\beta_r\,, \nonumber \\
\omega & = & \sum_r d\gamma_r \wedge d\beta_r\,.
\end{eqnarray}
The closure and nondegeneracy of $\omega$ are manifest, so it is
a symplectic two-form on ${\cal R}$; and the coordinates
$\beta, \gamma$ realize the local Darboux or canonical structure
for it. On the other hand, in these ``symplectic'' coordinates the
Fubini-Study metric is a bit involved. If we combine the
$\beta$'s and $\gamma$'s into a single column vector $\eta =
(\,\beta_1\, \beta_2\, \ldots\, \gamma_1\, \gamma_2\, \ldots\,)^T$, then the
length functional $L[C]$ of eq.(2.5) assumes the following local
form:
\begin{eqnarray}
L[C] & = & \int \!ds\, \sqrt{\dot{\eta}^T\, g(\eta)\, \dot{\eta}\,}\,,
\nonumber \\
g(\eta) & = &1\, +\, \frac{1}{2}\, \frac{\eta\, \eta^T}{1- {1 \over 2}\,\eta^T \, 
\eta } \,+\, \frac{1}{2} J\, \eta\, \eta^{T}\,J\,, \nonumber \\
J & = & \pmatrix{ 0 & 1 \cr -1 & 0}  \,,\;\;\; \eta^{T}\,\eta < 2\,. 
\end{eqnarray}
The symplectic matrix $J$ plays  a role in this expression for the
metric tensor matrix $g(\eta)$. This matrix $g(\eta)$ is checked  to be real 
symmetric positive definite, as one
eigenvalue is $(1-\frac{1}{2} \eta^T \,\eta)^{-1}$ (eigenvector
$\eta$), another eigenvalue is $(1-\frac{1}{2} \eta^T \,\eta)$
 (eigenvector $J\eta$), and the remaining eigenvalues are all
unity. We appreciate that for considerations of geometric phases
and null phase curves this kind of local description is really
appropriate, while free geodesics appear unavoidably complicated.

We also notice that, in case ${\cal H}$ is finite dimensional and the real 
dimension of the space ${\cal R}$ of unit rays is $2n$, the symplectic two-form 
$\omega$ of eq.(6.19) is invariant under the linear matrix group $Sp(2n,R)$ acting 
on the local coordinates $\beta, \gamma$. On the other hand, the integrand of the 
length functional $L[C]$ in eq.(6.20) possesses invariance only under 
$Sp(2n,R) \cap SO(2n) \simeq U(n)$, which is just the group of changes in the 
choice 
of the vectors $\{e_r\}$ which together with $\psi _0$ make up an orthonormal 
basis for ${\cal H}$.

Returning now to the discussions in Sections III and IV, we can
bring in submanifolds $M \subset {\cal R},\;\, {\cal M} = \pi^{-1} (M)
\subset {\cal H}$, with local coordinates $\xi^\mu, \alpha$ as
indicated in eqs. (4.1,2). Let $i_{\cal M}:\;{\cal  M} \hookrightarrow {\cal H}$
and $i_M : M \hookrightarrow {\cal R}$ be the corresponding
identification maps. Straightforward calculations show that the
pull-backs of $A, dA, \omega $ in eq.(6.15) to ${\cal M}$ and $M$ are locally
given (with mild abuse of notation) by:
\begin{eqnarray}
i^{\,*}_{\cal M}\, A & = & d\alpha + {\rm Im} \left(\psi(\xi ; \alpha), u_\mu(\xi ;
\alpha)\right) \,d\xi^\mu\,, \nonumber \\
i^{\,*}_{\cal M}\, dA \ = \ i^{\,*}_M \, \omega & = & {\rm Im} \left(u_\mu 
^\bot(\xi ; 
\alpha), u^\bot_\nu(\xi ; \alpha)\right)\, d\xi^\mu \wedge d\xi^\nu \nonumber \\
& = & {\rm Im} \left(u_\mu (\xi ; \alpha), u_\nu(\xi ; \alpha)\right)\, d\xi^\mu 
\wedge d\xi^\nu \,. 
\end{eqnarray}
We see, as is well known, that while the real symmetric part of
the hermitian matrix $\left((u^\bot_\mu , u^\bot_\nu)\right)$ determines the
metric, eq.(4.7), the imaginary antisymmetric part of the same
matrix is relevant for symplectic structure and geometric phase,
reinforcing the link between the latter two. (In case
$M={\cal R}$ and ${\cal M} = {\cal H}$, the $\xi^\mu$'s become the
$\beta$'s and $\gamma$'s of eqs. (6.17), and we immediately
recover the expressions (6.19)). For our present purposes the
following comments are pertinent: While $\omega$ is closed and
nondegenerate, $i^{\,*}_M \,\omega$ is closed but may well be degenerate. An
extreme case is when $M$ is an {\it isotropic submanifold} in
${\cal R}$, for then $i^{\,*}_M \,\omega = 0$. Such a situation can easily
arise if, for example, $M$ is described by a family of real
Schr\"{o}dinger wavefunctions $\psi(\xi ; q)$. (A Lagrangian
submanifold in ${\cal R}$ is a particular case of an isotropic
submanifold when the dimension is maximal, namely half the real
dimension of ${\cal R}$). One  may expect that if $M$ is isotropic
and $C \subset M$, then $C$ is a null phase curve. However this
 need not always be so, and the situation  is as follows.
For a general open curve $C_{12}$ from $\rho _1$ to $\rho _2$ in 
a general submanifold 
$M$, if we can find a null phase curve $C_{21}$ from $\rho _2$ to $\rho _1$  also 
lying in $M$, then $C_{12} \cup C_{21}$ is a closed loop; if $\pi _1 (M) = 0$, we can   
find a two-dimensional surface $S \in M$ having $C_{12} \cup C_{21}$ as boundary.  
 Then from eqn.(6.12) we obtain under these circumstances
\begin{eqnarray}
\varphi _g[C_{12}] = \varphi_g[C_{12} \cup C_{21}] = 
\int _{S \in M} i_M^{\;*}\,\omega\,.    \end{eqnarray}
\noindent
 Here as stated above, we had to choose $C_{21}$ to be a null phase curve. 
(In case $\rho_2 = \rho_1 $ and $C_{12}$ is already 
a closed loop, there is no need for any $C_{21}$;
it can be chosen to be trivial!) If however $M$ is an isotropic 
submanifold, { \it i.e.} $i_M^{\;*} \, \omega = 0$ 
(and assuming also $\pi_1(M)=0$), we can extract some very interesting consequences for geometric phases,
though it falls short of the vanishing of $\varphi_g[C]\/$ for every $C \subset M$.
We have the chain of implications
\begin{eqnarray}
i_M^{\;*}\omega =0 &\Leftrightarrow& \int_S i_M^{\;*}\omega =0, \mbox{
any two dimesional $S \subset M$}
\nonumber \\
&\Leftrightarrow & \varphi_g[C_{12} \cup C_{21}] =0,\;\; {\rm any} \rho_1, \rho_2
,C_{12},C_{21}\;\; {\rm in }\;\; M
\nonumber \\
&\Leftrightarrow&
\varphi_g[C_{12}]\;\; \mbox{unchanged under any continuous deformation}
\nonumber \\
&&\mbox{of $C_{12}$ leaving the end points $\rho_1, \rho_2$ fixed}
\end{eqnarray}
Thus, within an isotropic submanifold, the geometric phase for a general curve
depends on the two end points alone. In case the curve chosen is closed, it can be 
continuously shrunk to a point ( since, $\pi_1(M)=0\/$) and then its geometric phase 
vanishes. One can thus say in summary:
\begin{eqnarray}
M \subset {\cal R}, i_M^{\;.*} \omega &=&0,\quad \pi_1(M)=0, \quad C\subset M:
\nonumber \\
\varphi_g[C] &=&0 \;\; {\rm if} \quad \partial C=0;
\nonumber \\
\varphi_g[C]&=&\mbox{function of $\partial C\/$ alone, if $\partial C \neq 0$}.
\end{eqnarray}
The main conclusion is that general open curves in an isotropic submanifold 
need not be null phase curves; but geometric phases are invariant under continuous changes 
of their arguments leaving the end points unchanged. Perhaps this is not too surprising 
after all, since the isotropic property is a two-form condition.
\section{Concluding remarks}
We have shown that the familiar connection between the Bargmann invariants and 
geometric phases in quantum mechanics, based on the properties of free geodesics 
in ray and Hilbert spaces, can be generalised to a very significant extent.  In 
fact we have shown that our generalisation is the widest possible one.  The 
essential new concept is that of null phase curves in Hilbert and ray spaces -- the 
replacement of free geodesics by such curves leads to our generalisation.  We have 
seen through examples that 
 this wider connection between Bargmann invariants and geometric phases is just 
what is needed in several physically relevant situations.

Motivated by the fact that free geodesics are always null phase curves, we have 
defined the concept of constrained geodesics and posed the problem of determining 
when these may be null phase curves.  We have presented two examples when this is 
indeed so, 
 and one where they are not the same.  This reemphasizes the fact that constrained 
 geodesics and null phase curves are in principle different objects, and sharpens 
 the question of finding useful characterizations of the former which may ensure 
 the latter property for them.  This is sure to shed more light on the general 
questions raised in this paper, and we plan to return to them elsewhere.

\noindent
{\bf Acknowledgement:}     
One of us (EMR) thanks the Third World Academy of Sciences, Trieste, Italy for a 
 Travel Fellowship, the JNCASR, Bangalore, India for support as a Visiting 
 Scientist during June-August 1998, and the Centre for Theoretical Studies, IISc., 
 Bangalore for providing facilities during the completion of this work.


\begin{references}
\bibitem{berry}
M. V. Berry, Proc. Roy. Soc. London A {\bf 392}, 45(1984).
\bibitem{reprint}
Many of the early papers on geometric phase have been reprinted in A. Shapere and 
F. Wilczek, eds., Geometric Phases in Physics (World Scientific, Singapore, 1989); 
and in G. S. Agarwal, ed., Fundamentals of Quantum Optics (SPIE Milestone Series, 
SPIE Press, Bellington, 1995).
\bibitem{anandan}
Y. Aharonov and J. Anandan, Phys. Rev. Lett. {\bf 58}, 1593(1987).
\bibitem{samuel}
J. Samuel and R. Bhandari, Phys. Rev. Lett. {\bf 60}, 2339(1988).
\bibitem{nonabelian}
F. Wilczek and A. Zee, Phys. Rev. Lett. {\bf 52}, 2111(1984).
\bibitem{kinematic}
N. Mukunda and R. Simon, Ann. Phys. {\bf 228}, 205(1993); {\bf 228}, 269(1993).
\bibitem{hamilton}
W. R. Hamilton, {\it Lectures on Quaternions} (Dublin, 1853); L. C. Biedenharn and 
J. D. Louck, {\it Angular Momentum in Quantum Physics. Encyclopedia of Mathematics 
and its applications}, Vol. 8 (Addison-Wesley, Reading, MA, 1981). Hamilton's 
theory of turns has been generalized to the simplest noncompact semisimple group 
$SU(1,1) \sim SL(2,R)$ in R. Simon, N. Mukunda, and E.C.G. Sudarshan, Phys. Rev. 
Lett. {\bf 62}, 1331(1989); Jour. Math.  Phys. {\bf 30}, 1000(1989);
{\it Hamiltons turns for the Lorent group}, S. Chaturvedi, V.~Srinivasan, 
R. Simon and N. Mukunda,
{\it imsc preprint}
\bibitem{jphysa}
R. Simon and N. Mukunda. J. Phys. A: Math. Gen. {\bf 25}, 6135(1992).
\bibitem{bargmann}
V. Bargmann, J. Math. Phys. {\bf 5}, 862(1964).
\bibitem{wigner}
E. P. Wigner, {\it Group Theory} (Academic Press Inc., NY, 1959); 
J. Samuel, Pramana {\bf 48}, 959(1997).
\bibitem{page}
D. Page, Phys. Rev. {\bf A 36} , 3479(1987). 
\bibitem{kobayashi}
S. Kobayashi and K. Nomizu, {\it Foundations of Differential Geometry}, Vol. II 
(Interscience Publishers,  NY, 1969), Chap. IX.
\bibitem{pancharatnam}
S. Pancharatnam, Proc. Ind. Acad. Sci. Sect. A {\bf 44}, 247(1956). See also, S. 
Ramaseshan and R. Nityananda, Curr. Sci. {\bf 55}, 1225(1986); M. V. Berry, Jour. 
Mod. Opt. {\bf 34}, 1401(1987).
\bibitem{khanna}
G. Khanna, S. Mukhopadhyay, R. Simon, and N.  Mukunda, Ann. Phys. {\bf 253}, 
55(1997); Arvind, K. S. Mallesh, and N. Mukunda, J. Phys. A: Math. Gen. {\bf 30}, 
2417(1997). 
\bibitem{scvs}
S. Chaturvedi, M. S. Sriram, and V. Srinivasn, J. Phys. A: Math. Gen. {\bf 20}, 
{\bf L}1071(1987).
\bibitem{loba}
M. Berger, {\it Geometry II} (Springer-Verlag, Berlin, 1987), Chap.19; G. A. Jones 
and D. Singerman, {\it Complex Functions: An Algebraic and Geometric Viewpoint} 
(Cambridge University Press, Cambridge, 1987), Chap.5.
\bibitem{gouyprl}
R. Simon and N. Mukunda, Phys. Rev. Lett. {\bf 70}, 880(1993). 
\bibitem{gouy}
G. Gouy, C. R. Acad. Sci. Paris {\bf 110}, 125(1890); A. E. Siegman, {\it Lasers} 
(Oxford University Press, Oxford, 1986), Chap. 17.
\bibitem{simon}
B. Simon, Phys. Rev. Lett. {\bf 53}, 2167(1983).
\end{references}
\end{document}